\documentclass[aps, prd,  showpacs ]{revtex4}

\usepackage{amsmath,amsfonts,amscd,amssymb,mathrsfs,epsf,color}
\usepackage[small]{caption}

\newcommand{\pa}{\partial}

\newcommand{\vep}{\varepsilon}

\begin{document}

 \title{ Casimir interaction of concentric spheres at finite temperature}
\author{L. P. Teo}\email{ LeePeng.Teo@nottingham.edu.my}
\address{Department of Applied Mathematics, Faculty of Engineering, University of Nottingham Malaysia Campus, Jalan Broga, 43500, Semenyih, Selangor Darul Ehsan, Malysia.}

\begin{abstract}
We consider the finite temperature Casimir effect between two concentric spheres due to the vacuum fluctuations of the electromagnetic field in the $(D+1)$-dimensional Minkowski spacetime. Different combinations of perfectly conducting and infinitely permeable boundary conditions are imposed on the spheres. The asymptotic expansions of the Casimir free energies when the dimensionless parameter $\vep$, the ratio of the distance between the spheres to the radius of the smaller sphere, is small are derived in both the high temperature region and the low temperature region. It is shown that the leading terms     agree with   those obtained using the proximity force approximation, which are of order $T\vep^{1-D}$ in the high temperature region  and   of order $\vep^{-D}$ in the low temperature region. Some universal structures are observed in the next two correction terms.  The leading terms of the thermal corrections in the low temperature region are also  derived. They are found to be finite when $\vep\rightarrow 0^+$, and are of order $T^{D+1}$.

\end{abstract}

 \maketitle

\section{Introduction}
Casimir effect is one of the most interesting macroscopic phenomena in the quantum theory of fields. It has been under active studies under various context \cite{1}. The success in its experimental verification \cite{47,48,43,44,45,46} has intensified the interest in this effect. In recent years, the thermal correction to the Casimir effect has attracted increasing interest both theoretically and experimentally \cite{1,46,49,50,51}.

 The interest in the Casimir effect of spherical objects can be dated back to the work of Boyer \cite{2}, where he computed the zero temperature  Casimir force acting on a three-dimensional perfectly conducting spherical shell and found that it is repulsive. This result has later been confirmed in a number of other works \cite{3,4,5,6,7,8,9,10}. Since then, the Casimir effect in spherical configuration has attracted considerable interest. The cases of scalar fields, spinor fields and vector fields have been considered in various works \cite{11,12,13,14,15,16,17,27,20}. In fact, in \cite{12,13,14,15,16,17,20}, the authors considered spherical shells in general $(D+1)$-dimensional Minkowski spacetime rather than restricted to 4-dimensions. One of the motivations for this is that physics in higher-dimensional spacetimes   have become a trend since it was found that the existence of extra dimensions might be used to resolve some fundamental problems in physics such as the hierarchy problem. Another reason is that the dimension of spacetime can be used as a perturbation parameter in quantum field theory calculations \cite{12,18,19}.

For the last ten years, there has been an intense interest in studying the Casimir interaction between two objects. Several methods have been developed to compute the Casimir interaction   beyond the proximity force approximation, such as the    functional determinant or the multiple scattering method \cite{22,23,24,25,28,63} and the worldline approach \cite{53,29,52}. The corrections to the proximity force approximations have been computed for several geometric configurations such as the sphere-plane \cite{22,29,54,55,75,76}, cylinder-plane \cite{22,29,57,58,56}, cylinder-cylinder \cite{56,59,60}, sphere-sphere \cite{31,21}, etc. Recently, there has been an interest in considering the finite temperature correction to the Casimir interactions \cite{42,67,69,61,62}.

As a matter of fact, the  Casimir interaction between two concentric   spheres in $D=3$ dimensions has been considered in \cite{30,40,26}. For  scalar fields or spinor fields in general $D$-dimensions, the zero temperature Casimir effect on two concentric spherical shells  has been studied in \cite{39,64,65,66,70,71}.  In \cite{38}, we have derived  the zero temperature Casimir interaction between two  concentric spheres due to the fluctuations of electromagnetic field in the $D$-dimensional space. In this article,
we consider  the finite temperature effect. Moreover, we would derive the asymptotic behaviors of the Casimir free energy when the separation between the spheres is small.  The results are compared to the proximity force approximations.

In this article, we use units where $\hbar=c=k_B=1$.

\section{  Casimir free energy of concentric spheres}

Consider two concentric   spheres with radii $a_1$ and $a_2$ ($a_1<a_2$) in $(D+1)$-dimensional Minkowski spacetime, with either the perfectly conducting  or the infinitely permeable boundary conditions.
The electromagnetic field $F_{\mu\nu}=\pa_{\mu}A_{\nu}-\pa_{\nu}A_{\mu}$ satisfies the field equation:
\begin{equation*}
\frac{1}{\sqrt{|g|}}\pa_{\mu}\left( \sqrt{|g|} g^{\mu\kappa}g^{\nu\lambda}(\pa_{ \lambda}A_{\kappa}-\pa_{\kappa}A_{\lambda})\right)=0.
\end{equation*} As usual,  the Coulomb gauge $$\frac{1}{\sqrt{|g|}}\pa_{\mu}\left( \sqrt{|g|} A^{\mu}\right)=0$$is imposed to remove the gauge degree of freedom. The eigenmodes of the field are divided into TE modes and TM modes. In terms of $D$-dimensional spherical coordinates $(r,\boldsymbol{\theta}), \boldsymbol{\theta}=(\theta_1,\ldots,\theta_{D-2},\phi)$,
the TM modes have the form
\begin{equation*}
\begin{split}
A_r=& e^{-i\omega t}r^{-\frac{D}{2}}\left(C_1J_{\nu}(r\omega)+C_2N_{\nu}(r\omega)\right)Y_{l,\boldsymbol{m}}(\boldsymbol{\theta}),\\
\boldsymbol{A_{\theta}}=&e^{-i\omega t}r^{3-D}\frac{d}{dr}\left(r^{\frac{D-2}{2}}\left[C_1J_{\nu}(r\omega)+C_2N_{\nu}(r\omega)\right]\right)\boldsymbol{Y}_{l,\boldsymbol{m}}(\boldsymbol{\theta}).
\end{split}
\end{equation*}  The TE modes can be divided into $(D-2)$ sets,  each of them has the form
\begin{equation*}\begin{split}
A_r=& 0,\\
\boldsymbol{A_{\theta}}=&e^{-i\omega t} r^{\frac{4-D}{2}}\left(C_1J_{\nu}(r\omega)+C_2N_{\nu}(r\omega) \right)\boldsymbol{Y}_{l,\boldsymbol{m}}(\boldsymbol{\theta}).
\end{split}
\end{equation*}Here $\displaystyle \nu=l+\frac{D-2}{2}$, $\boldsymbol{m}=(m_2,\ldots,m_{D-1})$, $J_{\nu}(z)$ and $N_{\nu}(z)$ are Bessel functions of first kind and second kind. The sets of $\boldsymbol{m}$ have been discussed in detail in \cite{38}. For TM modes, each fixed $l$ has
$$b_l(D)=\frac{(2l+D-2)(l+D-3)!}{(D-2)! l!}$$allowable $\boldsymbol{m}$; whereas for TE modes, each fixed $l$ has
$$h_l(D)=\frac{l(l+D-2)(2l+D-2)(l+D-4)!}{(D-3)!(l+1)!}$$ allowable $\boldsymbol{m}$. The perfectly conducting boundary condition is equivalent to
$$\left.\left(\frac{\pa A_{\theta_i}}{\pa \theta_j}-\frac{\pa A_{\theta_j}}{\pa \theta_i}\right)\right|_{\text{boundary}}=0,\hspace{1cm} 1\leq i<j\leq D-1.$$
Therefore, for TE modes, the perfectly conducting boundary condition on $r=a_i$ implies
$$C_1J_{\nu}(a_i\omega)+C_2N_{\nu}(a_i\omega)=0,$$whereas for TM modes, we have
$$\frac{d}{dr}\left(r^{\frac{D-2}{2}}\left[C_1J_{\nu}(r\omega)+C_2N_{\nu}(r\omega)\right]\right)\Biggr|_{r=a_i}=0.$$
The infinitely permeable boundary condition is equivalent to
$$\left.\left(\frac{\pa A_{\theta_i}}{\pa r}-\frac{\pa A_{r}}{\pa \theta_i}\right)\right|_{\text{boundary}}=0,\hspace{1cm} 1\leq i\leq D-1.$$
Therefore, for TE modes the infinitely permeable boundary condition on $r=a_i$ implies
$$\frac{d}{dr}\left(r^{\frac{4-D}{2}}\left[C_1J_{\nu}(r\omega)+C_2N_{\nu}(r\omega)\right]\right)\Biggr|_{r=a_i}=0,$$whereas for TM modes, we have
$$C_1J_{\nu}(a_i\omega)+C_2N_{\nu}(a_i\omega)=0.$$

The interacting Casimir free energy $E_{\text{Cas}}$ of the concentric spheres can be written as a sum of the contribution from the TE and the TM modes:
\begin{equation}\label{eq6_13_1}E_{\text{Cas}}=E_{\text{TE}}+E_{\text{TM}}.\end{equation}In the following, we will use $E$ to represent either $E_{\text{TE}}$ or $E_{\text{TM}}$.
Using zeta regularization method, the Casimir free energy is given by (see e.g. \cite{1}, eq. (5.17)):
\begin{equation}\label{eq6_13_2}
E_{\text{Cas}}= -\frac{T}{2}\left(\zeta_T^{ \prime}(0)+\log[\mu^2]\zeta_T (0)\right),
\end{equation}where $\mu$ is a regularization parameter with the dimension of mass, $\zeta_T(s)$ is the zeta function
\begin{equation*}\begin{split}
\zeta_T (s)=&\sum_{\omega }\sum_{p=-\infty}^{\infty}\left(\omega^2+\xi_p^2\right)^{-s}, \end{split}
\end{equation*}$\omega$ are the TE or TM eigenfrequencies, and $\xi_p=2\pi p T$ are the Matsubara frequencies. As in the zero temperature case \cite{38}, one can show that
\begin{equation}\label{eq6_13_3}
\zeta_T(s)=\frac{2\sin\pi s}{\pi}\sum_{l=1}^{\infty}d_l(D)\sum_{p=0}^{\infty}\!' \int_{\xi_p}^{\infty}(\xi^2-\xi_p^2)^{-s}\frac{d}{d\xi}f_l(\xi)d\xi,
\end{equation} where $d_l(D)=h_l(D)$ for TE modes and $d_l(D)=b_l(D)$ for TM modes, and
\begin{equation}\label{eq5_6_5}f_{l}(\xi)=\ln\left(1-M_l(\xi)\right),\end{equation}
$$M_l(\xi)=\frac{\left(\alpha_1I_{\nu}(a_1\xi)+\beta_1a_1\xi I_{\nu}'(a_1\xi)\right)\left(\alpha_2K_{\nu}(a_2\xi)+\beta_2a_2\xi K_{\nu}'(a_2\xi)\right)}{\left(\alpha_2I_{\nu}(a_2\xi)+\beta_2a_2\xi I_{\nu}'(a_2\xi)\right)\left(\alpha_1K_{\nu}(a_1\xi)+\beta_1a_1\xi K_{\nu}'(a_1\xi)\right)},\quad l=1,2,\ldots.
$$ $I_{\nu}(z)$ and $K_{\nu}(z)$ are modified Bessel functions of first kind and second kind.
The values of $\alpha_i$ and $\beta_i$ depend on the type of modes and the boundary conditions imposed on the sphere $r=a_i$. They are listed in Table \ref{t1}.

\begin{table} [h]\caption{\label{t1}The values of $\alpha_i$ and $\beta_i$  under different   boundary conditions.}

\begin{tabular}{| c|p{0.3cm} c p{0.3cm}  |p{0.3cm} c p{0.3cm}  c p{0.3cm}   |}
\hline
\hline
Type of mode && Sphere $r=a_i$ && &$\alpha_i $&& $\beta_i$ &  \\
\hline
&&& &&&&& \\
TE && perfectly conducting &&   &1 && 0 & \\
&& &&& &&&\\
TM & &perfectly conducting   && & $\displaystyle \frac{D-2}{2}$ && 1  & \\
&&&&& &&&\\
TE && infinitely permeable  &&& $\displaystyle \frac{4-D}{2}$ && 1  & \\
&&&&& &&&\\
TM && infinitely permeable &&   &1 && 0 &  \\
&&&&&& &&\\
\hline
\hline
\end{tabular}

\end{table}

As in \cite{38}, one can show that
$$\sum_{l=1}^{\infty}d_l(D)\sum_{p=0}^{\infty}\!' \int_{\xi_p}^{\infty}(\xi^2-\xi_p^2)^{-s}\frac{d}{d\xi}f_l(\xi)d\xi$$ is an analytic function of $s$. Therefore,
$\zeta_T(0)=0$ and
\begin{equation*}
\zeta_T'(0)=2\sum_{l=1}^{\infty}d_l(D)\sum_{p=0}^{\infty}\!' \int_{\xi_p}^{\infty} \frac{d}{d\xi}f_l(\xi)d\xi=-2 \sum_{l=1}^{\infty}d_l(D)\sum_{p=0}^{\infty}\!' f_l(\xi_p).
\end{equation*}
Since
\begin{equation*}
I_{\nu}(z)=\left(\frac{z}{2}\right)^{\nu}\frac{1}{\Gamma(\nu+1)}\left[1+O(z^2)\right],\hspace{1cm}K_{\nu}(z)
=\left(\frac{z}{2}\right)^{-\nu} \frac{\Gamma(\nu )}{2} \left[1+O(z^2)\right],
\end{equation*}as $z\rightarrow 0$, we find that
\begin{equation*}
f_l(0)=\ln \left(1-\frac{(\alpha_1+\beta_1\nu)(\alpha_2-\beta_2\nu)}{(\alpha_1-\beta_1\nu)(\alpha_2+\beta_2\nu)}\left[\frac{a_1}{a_2}\right]^{2\nu}\right).
\end{equation*}
Therefore, from \eqref{eq6_13_2}, we find that the TE or TM contribution to the Casimir free energy is given by
\begin{equation}\label{eq4_21_2}
\begin{split}
E=&T\sum_{l=1}^{\infty}d_l(D)\left(\frac{1}{2}f_l(0)+\sum_{p=1}^{\infty}f_l(\xi_p)\right)
\\=&\frac{T}{2}\sum_{l=1}^{\infty}d_l(D)\ln \left(1-\frac{(\alpha_1+\beta_1\nu)(\alpha_2-\beta_2\nu)}{(\alpha_1-\beta_1\nu)(\alpha_2+\beta_2\nu)}\left[\frac{a_1}{a_2}\right]^{2\nu}\right)
+ T\sum_{l=1}^{\infty}\sum_{p=1}^{\infty}d_l(D) f_l(\xi_p).
\end{split}\end{equation}

Using Poisson summation formula,   the Casimir free energy \eqref{eq4_21_2} can also be written as
\begin{equation}\label{eq4_21_3}
\begin{split}
E=E_{ 0}+\frac{1}{\pi}\sum_{l=1}^{\infty}\sum_{p=1}^{\infty}d_l(D)\int_0^{\infty}f_l(\xi)\cos\frac{p\xi}{T}d\xi,
\end{split}
\end{equation}where
\begin{equation}\label{eq5_5_1}
E_{ 0}=\frac{1}{2\pi}\sum_{l=1}^{\infty}d_l(D)\int_0^{\infty}f_l(\xi)d\xi
\end{equation}is the zero temperature Casimir energy (vacuum energy). The expression \eqref{eq4_21_2} is suitable for the study of the high temperature limits of the Casimir free energies, but for the low temperature limits, the expression \eqref{eq4_21_3} would be preferred.

From the expression \eqref{eq4_21_2} for the Casimir free energy, one can use the argument in \cite{38} to show that the  force acting on the spheres is always attractive when the two spheres have the same boundary conditions (homogeneous boundary conditions); and is repulsive when one of the spheres is perfectly conducting and the other is infinitely permeable (mixed boundary conditions).
\section{Proximity force approximation}
In this section, we discuss briefly the proximity force approximation of the Casimir free energy when the separation between the spheres is small compared to both radii of the spheres. Define a dimensionless parameter $$\vep=\frac{a_2-a_1}{a_1}=\frac{d}{a_1},$$where $d=a_2-a_1$ is the distance between the two  spheres.
In the following, we are going to study the asymptotic behaviors of the Casimir free energy when $\vep \ll 1$. We consider the following two regions:
\begin{enumerate}
\item[1.] Low temperature:\; $  dT\ll a_2T\ll 1$;
\item[2.] High temperature:\; $a_2T\gg dT\gg 1$.
\end{enumerate}

For a pair of infinite parallel  plates  in $(D+1)$-dimensional spacetime, the Casimir free energy density, as a function of the separation between the plates $d$, is given by \cite{41,68}:
\begin{equation}\label{eq5_12_12}\mathcal{E}_{\text{Cas}}^{\parallel}(d)=(D-1)\left(-\frac{\Gamma\left(\frac{D+1}{2}\right)}{2^{D+1}\pi^{\frac{D+1}{2}}}\zeta_R(D+1)\frac{1}{d^D}
+\ldots\right),\end{equation}
or
\begin{equation}\label{eq5_12_13} \mathcal{E}_{\text{Cas}}^{\parallel}(d)=(D-1)\left(-\frac{\Gamma\left(\frac{D}{2}\right)}{2^D\pi^{\frac{D}{2}}}\zeta_R(D)\frac{T}{d^{D-1}}
+  \ldots\right),\end{equation}if both the two plates are perfectly conducting or infinitely permeable. We refer to these as the homogeneous boundary conditions. \eqref{eq5_12_12} is the low temperature asymptotics, the leading term being the zero temperature term; and \eqref{eq5_12_13} is the high temperature asymptotics, the leading term being called the classical term. The factor $(D-1)$ is due to the $(D-1)$ polarizations of photons in $(D+1)$-dimensional spacetime, $(D-2)$ of them come from the TE modes and one of them comes from the TM modes.

In case of mixed boundary conditions, i.e., one plate is perfectly conducting and one plate is infinitely permeable, the corresponding Casimir free energy density is \begin{equation*} \mathcal{E}_{\text{Cas}}^{\parallel}(d)=(D-1)\left(\frac{\Gamma\left(\frac{D+1}{2}\right)}{2^{D+1}\pi^{\frac{D+1}{2}}}
(1-2^{-D})\zeta_R(D+1)\frac{1}{d^D}
 +\ldots\right),\end{equation*}
or
\begin{equation*} \mathcal{E}_{\text{Cas}}^{\parallel}(d)=(D-1)\left(\frac{\Gamma\left(\frac{D}{2}\right)}{2^D\pi^{\frac{D}{2}}}(1-2^{1-D})\zeta_R(D)\frac{T}{d^{D-1}}
+  \ldots\right).\end{equation*}

In case of two concentric spheres, proximity force approximation of the Casimir free energy is particularly simple. It is given by the product of the surface area of the sphere (either one) with the Casimir free energy density between two parallel plates.
Since the surface area of a $(D-1)$-dimensional sphere of radius $a_1$ is
$$\mathscr{A}_{S^{D-1}}=\frac{2\pi^{\frac{D}{2}}}{\Gamma\left(\frac{D}{2}\right)}a_1^{D-1},$$ proximity force theorem implies that   in the low temperature region, the proximity force approximation to the Casimir free energy between the spheres is
\begin{equation}\label{eq5_10_1}
E_{\text{Cas}}^{\text{PFA}}\sim E_{\text{Cas},0}^{\text{PFA}}=(D-1)\left(-\frac{\Gamma\left(\frac{D+1}{2}\right)}{2^{D+1}\pi^{\frac{D+1}{2}}}\zeta_R(D+1)\frac{2\pi^{\frac{D}{2}}}{\Gamma\left(\frac{D}{2}\right)}
a_1^{D-1}\frac{1}{d^D}\right)=-\frac{1}{ \sqrt{\pi} a_1}\frac{D-1}{2^{D}\vep^{D}} \zeta_R(D+1)\frac{\Gamma\left(\frac{D+1}{2}\right)}{\Gamma
\left(\frac{D}{2}\right)}
\end{equation}for homogeneous boundary conditions, and is
\begin{equation}\label{eq5_10_2}
E_{\text{Cas}}^{\text{PFA}} \sim E_{\text{Cas},0}^{\text{PFA}}=\frac{1}{ \sqrt{\pi} a_1}\frac{D-1}{2^{D}\vep^{D}}(1-2^{-D}) \zeta_R(D+1)\frac{\Gamma\left(\frac{D+1}{2}\right)}{\Gamma
\left(\frac{D}{2}\right)}
\end{equation}for mixed boundary conditions. Notice that these leading terms are of order $\vep^{-D}$, $D$ being the space dimension.

In the high temperature region, the proximity force approximation gives
 \begin{equation}\label{eq5_10_3}
E_{\text{Cas}}^{\text{PFA}}\sim E_{\text{Cas}}^{\text{PFA,cl}}= - \frac{D-1}{2^{D-1}\vep^{D-1}} \zeta_R(D)T
\end{equation}for homogeneous boundary conditions, and
 \begin{equation}\label{eq5_10_4}
E_{\text{Cas}}^{\text{PFA}}\sim E_{\text{Cas}}^{\text{PFA,cl}}= \frac{D-1}{2^{D-1}\vep^{D-1}}(1-2^{1-D}) \zeta_R(D)T
\end{equation}
for mixed boundary conditions. These are of order $T\vep^{1-D}$.

In the following, we use the exact formulas \eqref{eq4_21_2} and \eqref{eq4_21_3} to find the asymptotic expansions of the Casimir free energy and compare to the proximity force approximations \eqref{eq5_10_1}, \eqref{eq5_10_2}, \eqref{eq5_10_3}, \eqref{eq5_10_4}. We start with the high temperature asymptotic expansion because it is less technical.

\section{Small separation asymptotic expansions   of the Casimir  free energies in the high temperature region}
From \eqref{eq4_21_2}, it is easy to see that in the high temperature region where $a_2T\gg dT\gg 1$, the Casimir free energy   is dominated by the term (classical term):
\begin{equation}\label{eq5_12_14}
E^{\text{cl}}=\frac{T}{2}\sum_{l=1}^{\infty}d_l(D)\ln \left(1-\frac{(\alpha_1+\beta_1\nu)(\alpha_2-\beta_2\nu)}{(\alpha_1-\beta_1\nu)(\alpha_2+\beta_2\nu)}\left[\frac{a_1}{a_2}\right]^{2\nu}\right).
\end{equation}
In the following, we derive the asymptotic expansion of this term when $\vep\ll 1$. We consider the case of homogeneous boundary conditions and the case of mixed boundary conditions separately.

\subsection{Homogeneous boundary conditions}
In the case both spheres have the same boundary conditions,
we find that
the   high temperature limit (the classical term)  of the Casimir free energy \eqref{eq5_12_14} is the same and is given by
\begin{equation*}\begin{split}
E^{\text{cl}} =&\frac{T}{2}\sum_{l=1}^{\infty}d_l(D)\ln \left(1- \left[\frac{a_1}{a_2}\right]^{2\nu}\right)
=\frac{T}{2}\sum_{l=1}^{\infty}d_l(D)\ln \left(1- e^{-2\alpha\nu}\right),
\end{split}\end{equation*}
where
\begin{equation}\label{eq5_11_2}\alpha=-\log\frac{a_1}{a_2}=\log\left(1+\frac{d}{a_1}\right)=\sum_{i=1}^{\infty}(-1)^{i-1}\frac{\vep^i}{i}.\end{equation}
Since $h_l(D)$ and $b_l(D)$ can be expanded as
\begin{equation*}
h_l(D)=\sum_{j=0}^{D-2}x_{D;j}\nu^j,\hspace{1cm}b_l(D)=\sum_{j=1}^{D-2}y_{D;j}\nu^j,
\end{equation*}we can write $d_l(D)$ as $\displaystyle d_l(D)=\sum_{j=0}^{D-2}\varpi_{D;j}\nu^j$. Then
\begin{equation}\label{eq5_11_1}
\begin{split}
E^{\text{cl}}
= &-\frac{T}{2}\sum_{j=0}^{D-2}\varpi_{D;j}\sum_{l=1}^{\infty}
\sum_{k=1}^{\infty}\frac{1}{k}\nu^{j }e^{-2\alpha k\nu}\\
= &-\frac{T}{2}\sum_{j=1}^{D-1}\varpi_{D;j-1}\int_{c-i\infty}^{c+i\infty}\Gamma(z)(2\alpha)^{-z}\zeta_R(z+1)\zeta_H\left(z-j+1;\tfrac{D}{2}\right)dz \\
\sim &-\frac{T}{2}\sum_{j=1}^{D-1} \frac{\varpi_{D;j-1}}{ 2^{j}}\Gamma(j)\zeta_R(j+1)\frac{1}{\alpha^j} +O(\ln\alpha).
\end{split}
\end{equation}
We have used the inverse Mellin transform formula
\begin{equation}\label{eq5_5_2}e^{-u}=\frac{1}{2\pi i}\int_{c-i\infty}^{c+i\infty}\Gamma(z)u^{-z} dz,\end{equation}
and the fact that the Hurwitz zeta function $\displaystyle\zeta_H\left(s;c\right)=\sum_{n=0}^{\infty}(n+c)^{-s}$ has a single simple pole at $s=1$ with residue $1$.
In fact, the use of the inverse Mellin transform \eqref{eq5_5_2} and the residue theorem allows us to find   the full asymptotic series in $\alpha$ from the second line of \eqref{eq5_11_1}. The last line \eqref{eq5_11_1} gives the asymptotic series up to the term in $\ln\alpha$. Using \eqref{eq5_11_2}, we can rewrite this asymptotic expansion in terms of $\vep$.
Since
\begin{equation}\label{eq5_11_3}
\begin{split}
x_{D;D-2}=&\frac{2}{ (D-3)!},\quad x_{D;D-3}=0,\quad x_{D;D-4}=-\frac{D^2-6D+32}{12 (D-4)!},\\
y_{D;D-2}=&\frac{2}{ (D-2)!},\quad y_{D;D-3}=0,\quad y_{D;D-4}=-\frac{1}{12 (D-5)!},
\end{split}\end{equation}
we find that when $D\geq 4$, the first three leading terms of the TE and TM contributions to the classical term are given respectively by
\begin{equation*}\begin{split}
E^{\text{cl}}_{\text{TE}}\sim &E^{\text{PFA,cl}}_{\text{TE}}\left\{1+\vep \frac{D-1}{2}+\vep^2\frac{(3D-8)(D-1)}{24}-\vep^2\frac{D^2-6D+32}{6(D-2)}\frac{\zeta_R(D-2)}{\zeta_R(D)}\right\},
\\
E^{\text{cl}}_{\text{TM}}\sim &E^{\text{PFA,cl}}_{\text{TM}}\left\{1+\vep \frac{D-1}{2}+\vep^2\frac{(3D-8)(D-1)}{24}-\vep^2\frac{ D-4}{6 }\frac{\zeta_R(D-2)}{\zeta_R(D)}\right\},
\end{split}
\end{equation*}where $E^{\text{PFA,cl}}_{\text{TE}}$ and $E^{\text{PFA,cl}}_{\text{TM}}$ are respectively the proximity force approximations for the TE and TM contributions which are $(D-2)/(D-1)$ and $1/(D-1)$ times the total proximity force approximation \eqref{eq5_10_3}.
The first three leading terms of the  Casimir free energy is then given by
\begin{equation}\label{eq5_11_7}\begin{split}
E^{\text{cl}}_{\text{Cas}}\sim E^{\text{PFA,cl}}_{\text{Cas}}\left\{1+\vep \frac{D-1}{2}+\vep^2\frac{(3D-8)(D-1)}{24}-\vep^2\frac{ D^2-5D+28}{6(D-1) }\frac{\zeta_R(D-2)}{\zeta_R(D)}\right\}.
\end{split}
\end{equation}
When $D=3$, a more precise computation shows that the first three leading terms of the Casimir free energy is
\begin{equation*}
E^{\text{cl}}_{\text{Cas}}\sim E^{\text{PFA,cl}}_{\text{Cas}}\left\{1+\vep+\frac{11}{6\zeta_R(3)}\vep^2\ln\vep\right\}=-\frac{T}{2\vep^2}\zeta_R(3)\left\{1+\vep+\frac{11}{6\zeta_R(3)}\vep^2\ln\vep\right\}.
\end{equation*}
This has equal contributions from the TE and the TM modes.
From \eqref{eq5_11_7}, we see that proximity force approximation underestimates the Casimir free energy, and the underestimation is worse when the space dimension becomes larger.
\subsection{Mixed boundary conditions}
In the case $r=a_1$ is perfectly conducting and $r=a_2$ is infinitely permeable, we find that the classical terms of the TE and TM contributions to the Casimir free energy are given respectively by
\begin{equation*}
E^{\text{cl} }_{\text{TE}}=\frac{T}{2}\sum_{l=1}^{\infty}h_l(D)\ln \left(1+\frac{\nu-\frac{4-D}{2}}{\nu+\frac{4-D}{2}} \left[\frac{a_1}{a_2}\right]^{2\nu}\right),\quad
E^{\text{cl} }_{\text{TM}}=\frac{T}{2}\sum_{l=1}^{\infty}b_l(D)\ln \left(1+\frac{\nu+\frac{D-2}{2}}{\nu-\frac{D-2}{2}} \left[\frac{a_1}{a_2}\right]^{2\nu}\right).
\end{equation*}
In the case $r=a_1$ is infinitely permeable and $r=a_2$ is perfectly conducting, we have
\begin{equation*}
E^{\text{cl} }_{\text{TE}}=\frac{T}{2}\sum_{l=1}^{\infty}h_l(D)\ln \left(1+\frac{\nu+\frac{4-D}{2}}{\nu-\frac{4-D}{2}} \left[\frac{a_1}{a_2}\right]^{2\nu}\right),\quad
E^{\text{cl} }_{\text{TM}}=\frac{T}{2}\sum_{l=1}^{\infty}b_l(D)\ln \left(1+\frac{\nu-\frac{D-2}{2}}{\nu+\frac{D-2}{2}} \left[\frac{a_1}{a_2}\right]^{2\nu}\right).
\end{equation*}
Consider series of the form
\begin{equation*}
I= \sum_{l=1}^{\infty}d_l(D)\ln\left(1+\frac{\nu+\vartheta}{\nu-\vartheta}\left[\frac{a_1}{a_2}\right]^{2\nu}\right)=\sum_{j=0}^{D-2}\varpi_{D;j}\sum_{l=1}^{\infty}\nu^j
\ln\left(1+\frac{\nu+\vartheta}{\nu-\vartheta}e^{-2\alpha\nu}\right),
\end{equation*}which can be rewritten as the sum of two terms:
 \begin{equation}\label{eq5_11_4}
\begin{split}
I= & \sum_{j=0}^{D-2}\varpi_{D;j}\sum_{l=1}^{\infty}\nu^{j } \left(\ln(1+e^{-2\alpha\nu})+\ln\left(1+\frac{2\vartheta}{\nu-\vartheta}\frac{1}{e^{2\alpha\nu}+1}\right)\right)\\
 =&\sum_{j=1}^{D-1}\varpi_{D;j-1}\sum_{l=1}^{\infty}\nu^{j-1 }\sum_{k=1}^{\infty}\frac{(-1)^{k-1}}{k}e^{-2\alpha k\nu}
 +\sum_{j=1}^{D-1}\varpi_{D;j-1}\sum_{l=1}^{\infty}\nu^{j-1 }\sum_{k=1}^{\infty}\frac{(-1)^{k-1}}{k}\frac{(2\vartheta)^k}{(\nu-\vartheta)^k}\frac{1}{(e^{2\alpha \nu}+1)^k}=I_1+I_2.
\end{split}
\end{equation}For the $I_1$ term, we find as before
\begin{equation*}
\begin{split}
I_1
\sim & \sum_{j=1}^{D-1} \frac{\varpi_{D;j-1}}{ 2^{j}}\Gamma(j)(1-2^{-j})\zeta_R(j+1)\frac{1}{\alpha^j} +O(\ln\alpha),
\end{split}
\end{equation*}where we have used
$$\sum_{k=1}^{\infty}\frac{(-1)^{k-1}}{k^s}=(1-2^{1-s})\zeta_R(s).$$
For the $I_2$ term,    since
\begin{equation*}
\begin{split}
\sum_{l=1}^{\infty}\frac{\nu^{j-1 }}{(\nu-\vartheta)^k}\frac{1}{(e^{2\alpha \nu}+1)^k}\sim &\int_{1}^{\infty}\frac{\left(x+\frac{D-2}{2}\right)^{j-1}}{
\left(x+\frac{D-2}{2}-\vartheta\right)^{k}}\frac{1}{\left(e^{\alpha(2x+D-2)}+1\right)^k}dx\\
\sim &\frac{1}{(2\alpha)^{j-k}}\sum_{r=0}^{j-k-1}\begin{pmatrix}k+r-1\\r\end{pmatrix}(2\alpha\vartheta)^{r} \lambda_{j-k-r-1,k}+O(\ln\alpha),
\end{split}
\end{equation*}where
$$\lambda_{\mu,\nu}=\int_0^{\infty}\frac{u^{\mu}}{(e^u+1)^{\nu}}du,$$we find that
\begin{equation*}\begin{split}
I_2=&\sum_{j=1}^{D-1}\varpi_{D;j-1} \sum_{k=1}^{j-1}\frac{(-1)^{k-1}}{k}\frac{(2\vartheta)^k}{(2\alpha)^{j-k}}\sum_{r=0}^{j-k-1}\begin{pmatrix}k+r-1\\r\end{pmatrix}(2\alpha\vartheta)^{r} \lambda_{j-k-r-1,k}+O(\ln\alpha).
\end{split}\end{equation*}
From these, we can derive the leading terms of the Casimir free energy.

If \textit{the sphere $r=a_1$ is perfectly conducting and the sphere $r=a_2$ is infinitely permeable}, then $\displaystyle \vartheta=\frac{D-4}{2}$ and $\displaystyle \vartheta=\frac{D-2}{2}$ respectively for TE and TM modes.
When $D\geq 4$, we find that the first three leading terms of the TE and TM contributions to the classical term are given respectively by
\begin{equation} \label{eq5_11_8}\begin{split}
E^{\text{cl}}_{\text{TE}}\sim E^{\text{PFA,cl}}_{\text{TE}}&\Biggl\{1+\vep \frac{D-1}{2}+\vep \frac{2(D-4)}{D-2}\frac{2^{D}-8}{2^{D}-2}\frac{\zeta_R(D-2)}{\zeta_R(D)}+\vep^2\frac{(3D-8)(D-1)}{24}\\&+\vep^2\frac{5D^2 - 30D + 16}{6(D-2)}\frac{2^D-8}{2^D-2}\frac{\zeta_R(D-2)}{\zeta_R(D)}+\vep^2\frac{2(D-4)^2}{(D-2)(D-3)}\frac{2^D-32}{2^D-2}\frac{\zeta_R(D-4)}{\zeta_R(D)}\Biggr\},
\\
E^{\text{cl}}_{\text{TM}}\sim E^{\text{PFA,cl}}_{\text{TM}}&\Biggl\{1+\vep \frac{D-1}{2}+2\vep \frac{2^{D}-8}{2^{D}-2}\frac{\zeta_R(D-2)}{\zeta_R(D)}+\vep^2\frac{(3D-8)(D-1)}{24}\\
&+\vep^2\frac{ 5D-8}{6 }\frac{2^D-8}{2^D-2}\frac{\zeta_R(D-2)}{\zeta_R(D)}+\vep^2\frac{2(D-2)}{(D-3)}\frac{2^D-32}{2^D-2}\frac{\zeta_R(D-4)}{\zeta_R(D)}\Biggr\},
\end{split}
\end{equation}where $E^{\text{PFA,cl}}_{\text{TE}}$ and $E^{\text{PFA,cl}}_{\text{TM}}$ are respectively the proximity force approximations for the TE and TM contributions which are $(D-2)/(D-1)$ and $1/(D-1)$ times the total proximity force approximation \eqref{eq5_10_4}. The first three leading terms of the total Casimir free energy is thus
\begin{equation}\label{eq5_11_9}\begin{split}
E^{\text{cl}}_{\text{Cas}}\sim E^{\text{PFA,cl}}_{\text{Cas}}&\Biggl\{1+\vep \frac{D-1}{2}+\vep \frac{2(D-3)}{D-1}\frac{2^{D}-8}{2^{D}-2}\frac{\zeta_R(D-2)}{\zeta_R(D)}+\vep^2\frac{(3D-8)(D-1)}{24}\\&+\vep^2\frac{5D^2 - 25D + 8}{6(D-1)}\frac{2^D-8}{2^D-2}\frac{\zeta_R(D-2)}{\zeta_R(D)}+\vep^2\frac{2(D^2-7D+14)}{(D-1)(D-3)}\frac{2^D-32}{2^D-2}\frac{\zeta_R(D-4)}{\zeta_R(D)}\Biggr\}.
\end{split}
\end{equation}When $D=5$, the  term $(2^D-32)\zeta_R(D-4)$ in \eqref{eq5_11_8} and \eqref{eq5_11_9} is understood as
$$\lim_{D\rightarrow 5} (2^D-32)\zeta_R(D-4)=32\ln 2.$$
When $D=3$, a more detail computation gives
\begin{align}\label{eq5_11_5}
E^{\text{cl}}_{\text{TE}}\sim &E^{\text{PFA,cl}}_{\text{TE}}\Biggl\{1+\vep  -\frac{8}{3}\vep\ln 2-\frac{2}{3\zeta_R(3)}\ln\vep \Biggr\}=\frac{3T}{16\vep^2}\zeta_R(3)\Biggl\{1+\vep  -\frac{8}{3}\vep\ln 2-\frac{2}{3\zeta_R(3)}\ln\vep \Biggr\},
\\\label{eq5_11_6}
E^{\text{cl}}_{\text{TM}}\sim & E^{\text{PFA,cl}}_{\text{TM}}\Biggl\{1+\vep  +\frac{8}{3}\vep\ln 2-\frac{2}{3\zeta_R(3)}\ln\vep\Biggr\}
=\frac{3T}{16\vep^2}\zeta_R(3)\Biggl\{1+\vep  +\frac{8}{3}\vep\ln 2-\frac{2}{3\zeta_R(3)}\ln\vep \Biggr\}\\
E^{\text{cl}}_{\text{Cas}}\sim & E^{\text{PFA,cl}}_{\text{Cas}}\Biggl\{1+\vep   -\frac{2}{3\zeta_R(3)}\ln\vep\Biggr\}
=\frac{3T}{8\vep^2}\zeta_R(3)\Biggl\{1+\vep  -\frac{2}{3\zeta_R(3)}\ln\vep \Biggr\}\nonumber.
\end{align}

If \textit{the sphere $r=a_1$ is infinitely permeable and the sphere $r=a_2$ is perfectly conducting}, then $\displaystyle \vartheta=\frac{4-D}{2}$ and $\displaystyle \vartheta=\frac{2-D}{2}$ respectively for the TE and TM modes.
When $D\geq 4$, the first three leading terms of the TE and TM contributions to the classical term are given respectively by
\begin{equation*}\begin{split}
E^{\text{cl}}_{\text{TE}}\sim E^{\text{PFA,cl}}_{\text{TE}}&\Biggl\{1+\vep \frac{D-1}{2}-\vep \frac{2(D-4)}{D-2}\frac{2^{D}-8}{2^{D}-2}\frac{\zeta_R(D-2)}{\zeta_R(D)}+\vep^2\frac{(3D-8)(D-1)}{24}\\&-\vep^2\frac{7D^2 - 42D + 80}{6(D-2)}\frac{2^D-8}{2^D-2}\frac{\zeta_R(D-2)}{\zeta_R(D)}+\vep^2\frac{2(D-4)^2}{(D-2)(D-3)}\frac{2^D-32}{2^D-2}\frac{\zeta_R(D-4)}{\zeta_R(D)}\Biggr\},
\\
E^{\text{cl}}_{\text{TM}}\sim E^{\text{PFA,cl}}_{\text{TM}}&\Biggl\{1+\vep \frac{D-1}{2}-2\vep \frac{2^{D}-8}{2^{D}-2}\frac{\zeta_R(D-2)}{\zeta_R(D)}+\vep^2\frac{(3D-8)(D-1)}{24}\\
&-\vep^2\frac{ 7D-16}{6 }\frac{2^D-8}{2^D-2}\frac{\zeta_R(D-2)}{\zeta_R(D)}+\vep^2\frac{2(D-2)}{(D-3)}\frac{2^D-32}{2^D-2}\frac{\zeta_R(D-4)}{\zeta_R(D)}\Biggr\}.
\end{split}
\end{equation*} The first three leading terms of the total Casimir free energy is thus
\begin{equation*}\begin{split}
E^{\text{cl}}_{\text{Cas}}\sim E^{\text{PFA,cl}}_{\text{Cas}}&\Biggl\{1+\vep \frac{D-1}{2}-\vep \frac{2(D-3)}{D-1}\frac{2^{D}-8}{2^{D}-2}\frac{\zeta_R(D-2)}{\zeta_R(D)}+\vep^2\frac{(3D-8)(D-1)}{24}\\&+\vep^2\frac{7D^2 - 35D + 64}{6(D-1)}\frac{2^D-8}{2^D-2}\frac{\zeta_R(D-2)}{\zeta_R(D)}+\vep^2\frac{2(D^2-7D+14)}{(D-1)(D-3)}\frac{2^D-32}{2^D-2}\frac{\zeta_R(D-4)}{\zeta_R(D)}\Biggr\}.
\end{split}
\end{equation*}When $D=3$, by duality, the TE contribution is given by \eqref{eq5_11_6}, and the TM contribution is given by \eqref{eq5_11_5}.

Observe that the corrections to the proximity force approximation in the case of mixed boundary conditions is more complicated than the case of homogeneous boundary conditions. We also find that the proximity force approximations underestimate the Casimir free energies.

Compare the first correction to the proximity force approximation for the two scenarios of mixed boundary conditions, we find that in the high temperature region, the force is stronger when the sphere with smaller radius is perfectly conducting and the sphere with larger radius is infinitely permeable.
\section{Small separation asymptotic expansions   of the Casimir free energies in the low temperature region}\label{sl} In the low temperature region, the Casimir free energy is dominated by the zero temperature term.
Making a change of variables $\xi\mapsto \omega/a_1$ in \eqref{eq5_5_1}, we find that
the zero temperature Casimir free energy  can be written as
 \begin{equation*}
\begin{split}
E_{ 0}=&\frac{1}{2\pi a_1}\sum_{l=1}^{\infty}d_l(D)\int_0^{\infty}\ln (1-A_{\nu}(\omega))d\omega\\
=&-\frac{1}{2\pi a_1}\sum_{s=1}^{\infty}\frac{1}{s}\sum_{l=1}^{\infty}d_l(D)\nu\int_0^{\infty}A_{\nu}(\nu\omega)^sd\omega,
\end{split}\end{equation*}where
\begin{equation*}
A_{\nu}(\omega)=\frac{\left(\alpha_1I_{\nu}(\omega)+\beta_1\omega I_{\nu}'(\omega)\right)\left(\alpha_2K_{\nu}(\omega(1+\vep))+\beta_2\omega(1+\vep) K_{\nu}'(\omega(1+\vep))\right)}{\left(\alpha_1K_{\nu}(\omega)+\beta_1\omega K_{\nu}'(\omega)\right)\left(\alpha_2I_{\nu}(\omega(1+\vep))+\beta_2\omega(1+\vep)I_{\nu}'(\omega(1+\vep))\right)}.
\end{equation*}
From Debye asymptotic expansions of Bessel functions \cite{36,37}, we have
\begin{equation*}
\begin{split}
&\frac{ \alpha_iI_{\nu}(\nu\omega)+\beta_i\nu\omega I_{\nu}'(\nu\omega)  }{ \alpha_iK_{\nu}(\nu\omega)+\beta_i\nu\omega K_{\nu}'(\nu\omega) }
=\left\{\begin{aligned}&\frac{1}{\pi}\exp\left(2\nu\eta(\omega)+2\sum_{k=1}^{\infty}\frac{D_{2k-1}(t(\omega))}{\nu^{2k-1}}\right),\quad \text{if}\;\;\alpha_i=1,\beta_i=0\\
&-\frac{1}{\pi}\exp\left(2\nu\eta(\omega)+2\sum_{k=1}^{\infty}\frac{M_{2k-1,\alpha_i}(t(\omega))}{\nu^{2k-1}}\right),\quad \text{if}\;\; \beta_i=1\end{aligned}\right.,
\end{split}
\end{equation*}
where
\begin{equation*}\begin{split}\eta(z)=&\sqrt{1+z^2}+\log\frac{z}{1+\sqrt{1+z^2}},\hspace{1cm}t(z)=\frac{1}{\sqrt{1+z^2}},\\
\sum_{k=1}^{\infty}\frac{D_{k}(t)}{\nu^k}= &\ln\left(1+\sum_{k=1}^{\infty}\frac{ u_{k }(t)}{\nu^k}\right),\hspace{1cm}
\sum_{k=1}^{\infty}\frac{M_{k,\alpha}(t)}{\nu^k}= \ln\left(1+\sum_{k=1}^{\infty}\frac{v_k(t)+\alpha tu_{k-1}(t)}{\nu^k}\right),\end{split}\end{equation*}
$u_k(t)$ and $v_k(t)$ are polynomials in $t$ defined recursively by
\begin{align*}
&u_0(t)=1, \hspace{0.5cm}u_{k}(t)=\frac{t^2(1-t^2)}{2}u_{k-1}'(t)+\frac{1}{8}\int_0^t(1-5\tau^2)u_{k-1}(\tau)d\tau,\\
&v_0(t)=1,\hspace{0.5cm} v_{k }(t)=u_k(t)-t^2(1-t^2)u_{k-1}'(t)-\frac{t(1-t^2)}{2}u_{k-1}(t).
\end{align*}
In the following, we discuss the asymptotic expansions of the zero temperature Casimir free energy for the case of homogeneous boundary conditions and the case of mixed boundary conditions separately.

\subsection{Homogeneous boundary conditions}
In this case, $\alpha_1=\alpha_2,\beta_1=\beta_2$. We find that
\begin{equation}\label{eq5_11_11}
A_{\nu}(\nu\omega)\sim \exp\left(-2\nu\left(\eta([1+\vep]\omega)-\eta(\omega)\right)-2\sum_{k=1}^{\infty}\frac{P_{2k-1}(t([1+\vep]\omega))-P_{2k-1}(t(\omega))}{\nu^{2k-1}}\right).
\end{equation}The polynomials $P_k(t)$ are equal to $D_k(t)$ or $M_{k,\alpha}(t)$  depending on the type of modes and the boundary conditions, as shown in Table \ref{t2}.

\begin{table} [h]\caption{\label{t2}The polynomial $P_k(t)$ under different   boundary conditions.}

\begin{tabular}{| c|p{0.3cm} c p{0.3cm}  |p{0.3cm} c   p{0.3cm}   |}
\hline
\hline
Type of mode && Boundary conditions  on both spheres && & $P_k(t)$ &  \\
\hline
&&& &&& \\
TE && perfectly conducting &&   & $D_k(t)$ & \\
&& &&& & \\
TM & &perfectly conducting   && & $\displaystyle M_{k,\frac{D-2}{2}}(t)$ &  \\
&&&&& & \\
TE && infinitely permeable  &&& $\displaystyle M_{k,\frac{4-D}{2}}(t)$ &  \\
&&&&& & \\
TM && infinitely permeable &&   &$D_k(t)$ &   \\
&&&&&&  \\
\hline
\hline
\end{tabular}

\end{table}

In the following, we only find the first three terms in the asymptotic expansion when $\vep\rightarrow 0^+$. For this, it is sufficient to take the $k=1$ term in \eqref{eq5_11_11}. The polynomials $D_1(t)$ and $M_{1,\alpha}(t)$ are given explicitly by
\begin{equation*}
D_1(t)=\frac{t}{8}-\frac{5t^3}{24},   \hspace{1cm} M_{1,\alpha}(t)=\left(\alpha-\frac{3}{8}\right)t+\frac{7t^3}{24}.\end{equation*}Therefore, we can write $P_1(t)$ as
$P_1(t)=\lambda t +\gamma t^3.$
Using \eqref{eq5_5_2}, we have
\begin{equation*}
\begin{split}
E_0\sim&-\frac{1}{2\pi a_1}\sum_{s=1}^{\infty}\frac{1}{s}\sum_{j=0}^{D-2}\varpi_{D;j}\nu^{j+1}\int_0^{\infty}
\exp\left(-2s\left[\nu\left(\eta((1+\vep)\omega)-\eta(\omega)\right)+ \frac{P_{1}(t(\omega(1+\vep)))-P_{1}(t(\omega))}{\nu}\right]\right)
d\omega\\
\sim &-\frac{1}{2\pi a_1} \sum_{l=1}^{\infty}\sum_{j=0}^{D-2}\varpi_{D;j}\nu^{j+1}\int_0^{\infty}\frac{1}{2\pi i}\int_{c-i\infty}^{c+i\infty}\Gamma(z)2^{-z}\zeta_R(z+1)\nu^{-z}\left(\eta((1+\vep)\omega)-\eta(\omega)\right)^{-z}\\&\hspace{3cm}\times
\left(1+ \frac{1}{\nu^{2}}\frac{P_{1}(t(\omega(1+\vep)))-P_{1}(t(\omega))}{\eta((1+\vep)\omega)-\eta(\omega)} \right)^{-z}dzd\omega.
\end{split}
\end{equation*}
Since
\begin{equation*}\begin{split}
&\left(\eta((1+\vep)\omega)-\eta(\omega) \right)^{-z}=    \vep^{-z}\omega^{-z}\eta'(\omega)^{-z}\left(1-z\frac{\vep \omega }{2}\frac{\eta^{\prime\prime}(\omega)}{\eta'(\omega)}+\vep^2\left[- \frac{ z \omega^2}{6}\frac{\eta^{\prime\prime\prime}(\omega)}{\eta'(\omega)}+ \frac{  z(z+1)\omega^2 }{8}\frac{\eta^{\prime\prime}(\omega)^2}{\eta'(\omega)^2}\right]+\ldots\right),
\\
&\left(1+ \frac{1}{\nu^{2}}\frac{P_{1}(t(\omega(1+\vep)))-P_{1}(t(\omega))}{\eta((1+\vep)\omega)-\eta(\omega)} \right)^{-z}=1-\frac{z}{\nu^2}\frac{P_1'(t(\omega))}{\eta'(\omega)}t'(\omega)+\ldots,
\end{split}
\end{equation*}we find that
\begin{equation*}
\begin{split}
E_0\sim&-\frac{1}{2\pi a_1}  \sum_{j=0}^{D-2}\varpi_{D;j} \frac{1}{2\pi i}\int_{c-i\infty}^{c+i\infty}\Gamma(z)2^{-z}\zeta_R(z+1) \vep^{-z} \Bigl(\zeta_H\left(z-j-1;\tfrac{D}{2}\right)\mathcal{A}(z)-z\mathcal{B}(z)  \zeta_H\left(z-j+1;\tfrac{D}{2}\right) \Bigr) +\ldots,
\end{split}
\end{equation*}where
\begin{equation*}
\begin{split}
\mathcal{A}(z)=&\int_0^{\infty}(\omega\eta'(\omega))^{-z}\left(1-z\frac{\vep \omega }{2}\frac{\eta^{\prime\prime}(\omega)}{\eta'(\omega)}+\vep^2\left[- \frac{ z \omega^2}{6}\frac{\eta^{\prime\prime\prime}(\omega)}{\eta'(\omega)}+ \frac{ z(z+1)\omega^2 }{8}\frac{\eta^{\prime\prime}(\omega)^2}{\eta'(\omega)^2}\right]\right)
 d\omega,\\
 \mathcal{B}(z)=&\int_0^{\infty}(\omega\eta'(\omega))^{-z}\frac{P_1'(t(\omega))t'(\omega)}{\eta'(\omega)} d\omega.
\end{split}
\end{equation*}It is straightforward to find that
\begin{equation*}
\begin{split}
 \mathcal{A}(z)=& \frac{\sqrt{\pi}}{2}\frac{\Gamma\left(\frac{z-1}{2}\right)}{\Gamma\left( \frac{z}{2}\right)}\left(1+\vep\frac{z-1}{2}
 + \frac{\vep^2}{24}\frac{(z - 1)(3z^2 - 2z - 17)}{z+2}\right),
\\
\mathcal{B}(z)=&\frac{\sqrt{\pi}}{2}\frac{\Gamma\left(\frac{z+1}{2}\right)}{\Gamma\left( \frac{z+2}{2}\right)}\left(-\lambda+(\lambda-3\gamma)\frac{z+1}{z+2}+3\gamma\frac{(z+1)(z+3)}{(z+2)(z+4)}\right).
\end{split}
\end{equation*}From this, residue theorem gives
\begin{equation*}
\begin{split}
E_0\sim&
-\frac{1}{2\pi a_1}  \sum_{j=0}^{D-2}\frac{\varpi_{D;j}}{2^{j+2}\vep^{j+2}}  \Gamma(j+2) \zeta_R(j+3) \mathcal{A}(j+2)
+\frac{1}{2\pi a_1}  \sum_{j=1}^{D-2}\frac{\varpi_{D;j}}{2^{j}\vep^{j}}  \Gamma(j+1) \zeta_R(j+1) \mathcal{B}(j)+\ldots\\\sim & -\frac{1}{2\pi a_1}\frac{\varpi_{D;D-2}}{(2\vep)^D}\Gamma(D)\zeta_R(D+1)\mathcal{A}(D)-\frac{1}{2\pi a_1}\frac{\varpi_{D;D-4}}{(2\vep)^{D-2}}\Gamma(D-2)\zeta_R(D-1)\mathcal{A}(D-2)\\&+\frac{1}{2\pi a_1}\frac{\varpi_{D;D-2}}{(2\vep)^{D-2}}\Gamma(D-1)\zeta_R(D-1)\mathcal{B}(D-2).
\end{split}
\end{equation*}

For \textit{perfectly conducting boundary conditions on both spheres}, we have
$\displaystyle\lambda=\frac{1}{8},\ \gamma=-\frac{5}{24}$ for TE modes and
$\displaystyle\lambda=\frac{D-2}{2}-\frac{3}{8}, \gamma=\frac{7}{24}$ for TM modes. Using \eqref{eq5_11_3}, we find that if $D\geq 4$, the first three leading terms of the TE  and TM contributions to the zero temperature Casimir  free energy are given respectively by
\begin{equation*}
\begin{split}
E_{\text{TE},0}\sim  &E_{\text{TE},0}^{\text{PFA}}\Biggl\{1+\vep\frac{D-1}{2}+\vep^2\frac{(D-1)(3D^2-2D-17)}{24(D+2)} -\vep^2\frac{D^4-4D^3+20D^2+76D-21}{6 D(D-1)(D+2)}\frac{\zeta_R(D-1)}{\zeta_R(D+1)}\Biggr\},\\
E_{\text{TM},0}\sim  &E_{\text{TM},0}^{\text{PFA}}\Biggl\{1+\vep\frac{D-1}{2}+\vep^2\frac{(D-1)(3D^2-2D-17)}{24(D+2)} -\vep^2\frac{D^4 - 4D^3 - 16D^2 + 4D + 87}{6D(D+2)(D-1)}\frac{\zeta_R(D-1)}{\zeta_R(D+1)}\Biggr\},
\end{split}
\end{equation*}where $E^{\text{PFA}}_{\text{TE},0}$ and $E^{\text{PFA}}_{\text{TM},0}$ are respectively the proximity force approximations for the TE and TM contributions which are $(D-2)/(D-1)$ and $1/(D-1)$ times the total proximity force approximation \eqref{eq5_10_1}. When $D=3$, the term $\varpi_{D,D-4}$ has to be set to zero. One obtains
\begin{align}\label{eq5_11_12}
E_{\text{TE},0}\sim & -\frac{\pi^3}{360a_1\vep^3}\left(1+\vep+\frac{\vep^2}{15}-\vep^2\frac{5}{4\pi^2}\right),
\\\label{eq5_11_13}
E_{\text{TM},0}\sim & -\frac{\pi^3}{360a_1\vep^3}\left(1+\vep+\frac{\vep^2}{15}+\vep^2\frac{19}{4\pi^2}\right).
\end{align}
Summing the TE and TM contributions, we find that if $D\geq 4$, the asymptotic expansion of the zero temperature Casimir free energy is given by
\begin{equation*}
\begin{split}
E_{\text{Cas},0}\sim &E_{\text{Cas},0}^{\text{PFA}}\Biggl\{1+\vep\frac{D-1}{2}+\vep^2\frac{(D-1)(3D^2-2D-17)}{24(D+2)} -\vep^2\frac{D^4 - 4D^3 + 20D^2 + 40D - 129}{6D(D+2)(D-1)}\frac{\zeta_R(D-1)}{\zeta_R(D+1)}\Biggr\};
\end{split}
\end{equation*}and if $D=3$,
\begin{equation*}
E_{\text{Cas},0}\sim -\frac{\pi^3}{180a_1\vep^3}\left(1+\vep+\frac{\vep^2}{15}+\vep^2\frac{7}{4\pi^2}+\ldots\right).
\end{equation*}

For \textit{infinitely permeable boundary conditions on both spheres},
$\displaystyle \lambda=\frac{4-D}{2}-\frac{3}{8},  \gamma=\frac{7}{24}$ for TE modes and $\displaystyle\lambda=\frac{1}{8},\quad\gamma=-\frac{5}{24}$
for TM modes. Therefore, we find that if $D\geq 4$, the first three leading terms of the TE and TM contributions to the zero temperature Casimir free energy are given respectively by
\begin{equation*}
\begin{split}
E_{\text{TE},0}\sim  &E_{\text{TE},0}^{\text{PFA}}\Biggl\{1+\vep\frac{D-1}{2}+\vep^2\frac{(D-1)(3D^2-2D-17)}{24(D+2)}  -\vep^2\frac{D^3 - 3D^2 + 29D + 57}{6D(D+2) }\frac{\zeta_R(D-1)}{\zeta_R(D+1)}\Biggr\},
\\
E_{\text{TM},0}\sim  &E_{\text{TM},0}^{\text{PFA}}\Biggl\{1+\vep\frac{D-1}{2}+\vep^2\frac{(D-1)(3D^2-2D-17)}{24(D+2)} -\vep^2\frac{(D^2-7)(D-3)}{6 D(D+2)}\frac{\zeta_R(D-1)}{\zeta_R(D+1)}\Biggr\}.
\end{split}
\end{equation*} When $D=3$, the TE contribution is given by \eqref{eq5_11_13}, and the TM contribution is given by \eqref{eq5_11_12} due to duality.
Summing the TE and TM contributions, we find that if $D\geq 4$, the asymptotic expansion of the zero temperature Casimir free energy is given by
\begin{equation*}
\begin{split}
E_{\text{Cas},0}\sim &E_{\text{Cas},0}^{\text{PFA}}\Biggl\{1+\vep\frac{D-1}{2}+\vep^2\frac{(D-1)(3D^2-2D-17)}{24(D+2)}-\vep^2\frac{D^4 - 4D^3 + 32D^2 - 8D - 93}{6 D(D+2)(D-1)}\frac{\zeta_R(D-1)}{\zeta_R(D+1)}\Biggr\}.
\end{split}
\end{equation*}
It is interesting to note that in the case of homogeneous boundary conditions, the first analytic correction to the Casimir free energy has the form
$$E=E^{\text{PFA}}\left(1+\vep\frac{D-1}{2}+\ldots\right) $$ both in the high temperature region and the low temperature region.  This is already true    for the TE and TM contributions separately. It also follows that the proximity force approximation always underestimates the strength of the  force.
\subsection{Mixed boundary conditions}
In this case, $\alpha_1\neq \alpha_2,\beta_1\neq \beta_2$. The expression for $A_{\nu}(\nu(\omega))$ is more complicated:
\begin{equation*}
A_{\nu}(\nu\omega)\sim \exp\left(-2\nu\left(\eta([1+\vep]\omega)-\eta(\omega)\right)-2\sum_{k=1}^{\infty}\frac{Q_{2k-1}(t([1+\vep]\omega))-P_{2k-1}(t(\omega))}{\nu^{2k-1}}\right)
\end{equation*}The polynomials $P_k(t)$ and $Q_k(t)$ are equal to $D_k(t)$ or $M_{k,\alpha}(t)$  depending on the type of modes and the boundary conditions, as shown in Table \ref{t3}.

\begin{table} [h]\caption{\label{t3}The polynomial $P_k(t)$ and $Q_{k}$ under different   boundary conditions.}

\begin{tabular}{| c|p{0.3cm} c p{0.3cm}  |p{0.3cm} c p{0.3cm}  |p{0.3cm} c p{0.2cm}| p{0.2cm}  c p{0.3cm}   |}
\hline
\hline
Type of mode &&   Sphere 1 &&&   Sphere 2 && & $P_k(t)$ &&&   $Q_k(t)$ &  \\
\hline
&&& &&&&& &&&&\\
TE && perfectly conducting &&& infinitely permeable  && & $D_k(t)$ &&& $M_{k,\frac{4-D}{2}}(t)$ & \\
&& &&& &&&&&&&\\
TM & &perfectly conducting  &&& infinitely permeable  && & $\displaystyle M_{k,\frac{D-2}{2}}(t)$ &&& $\displaystyle D_k(t)$  & \\
&&&&& &&&&&&&\\
TE && infinitely permeable &&& perfectly conducting &&& $M_{k,\frac{4-D}{2}}(t)$ &&& $\displaystyle D_k(t)$  & \\
&&&&& &&&&&&&\\
TM && infinitely permeable &&& perfectly conducting &&   &$D_k(t)$ &&&$\displaystyle M_{k,\frac{D-2}{2}}(t)$ &  \\
&&&&&& &&&&&&\\
\hline
\hline
\end{tabular}

\end{table}

In the present case, the computation is more involved because $P_k(t)\neq Q_k(t)$. Proceed as in the previous section, we find that
\begin{equation}\label{eq5_6_1}
\begin{split}
E_0
\sim &\frac{1}{2\pi a_1}  \sum_{j=0}^{D-2}\varpi_{D;j} \frac{1}{2\pi i}\int_{c-i\infty}^{c+i\infty}\Gamma(z)2^{-z}(1-2^{-z})\zeta_R(z+1) \vep^{-z} \Biggl(\zeta_H\left(z-j-1;\tfrac{D}{2}\right)\mathcal{A}(z)-z\mathcal{C}(z)  \zeta_H\left(z-j+1;\tfrac{D}{2}\right)\\&\hspace{3cm}+\frac{1}{\vep^2}\frac{z(z+1)}{2}\mathcal{G}(z) \zeta_H\left(z-j+3;\tfrac{D}{2}\right) \Biggr) +\ldots,
\end{split}
\end{equation}where $\mathcal{A}(z)$ is the same as before,
\begin{equation*}
\begin{split}
 \mathcal{C}(z)=&\int_0^{\infty}(\omega\eta'(\omega))^{-z}\left[\frac{Q_1'(t(\omega))t'(\omega)}{\eta'(\omega)} -(z+1)\frac{ \mathcal{T}(t(\omega))}{2}\frac{\eta^{\prime\prime}(\omega)}{\eta^{\prime}(\omega)^2}+\frac{1}{\vep}\frac{\mathcal{T}(t(\omega))}{\omega\eta'(\omega)}\right]d\omega,\\
  \mathcal{G}(z)=&\int_0^{\infty}(\omega\eta'(\omega))^{-z-2}\mathcal{T}(t(\omega))^2d\omega.
\end{split}
\end{equation*}
Here $\mathcal{T}(t)=Q_1(t)-P_1(t)$ can be written as $\mathcal{T}(t)=\delta t+\kappa t^3$. On the other hand, write $Q_1(t)$ as $Q_1(t)=\lambda t+\gamma t^3$,
we find that
\begin{equation*}
\begin{split}
\mathcal{C}(z)=&\frac{\sqrt{\pi}}{2}\frac{\Gamma\left(\frac{z+1}{2}\right)}{\Gamma\left( \frac{z+2}{2}\right)}\left( -\lambda+\left(\lambda-3\gamma+\frac{z+1}{2}\delta\right)\frac{z+1}{z+2}+\left(3\gamma+\frac{z+1}{2}\kappa\right)\frac{(z+1)(z+3)}{(z+2)(z+4)}
+\frac{1}{\vep}\left[\delta+\kappa\frac{z+1}{z+2}\right]\right),\\
\mathcal{G}(z)=& \frac{\sqrt{\pi}}{2}\frac{\Gamma\left(\frac{z+3}{2}\right)}{\Gamma\left(\frac{z+4}{2}\right)}\left(\delta^2+2\delta\kappa \frac{z+3}{z+4}+\kappa^2\frac{(z+3)(z+5)}{(z+4)(z+6)}\right).
\end{split}
\end{equation*}
Then
\begin{equation*}
\begin{split}
E_0\sim&
\frac{1}{2\pi a_1}  \sum_{j=0}^{D-2}\frac{\varpi_{D;j}}{2^{j+2}\vep^{j+2}}  \Gamma(j+2) (1-2^{-j-2})\zeta_R(j+3) \mathcal{A}(j+2)
-\frac{1}{2\pi a_1}  \sum_{j=1}^{D-2}\frac{\varpi_{D;j}}{2^{j}\vep^{j}}  \Gamma(j+1) (1-2^{-j})\zeta_R(j+1) \mathcal{C}(j)
\\&+\frac{1}{2\pi a_1}  \sum_{j=3}^{D-2}\frac{\varpi_{D;j}}{2^{j-1}\vep^{j }}  \Gamma(j) (1-2^{2-j})\zeta_R(j-1) \mathcal{G}(j-2)+\ldots\\
\sim&
\frac{1}{2\pi a_1}   \frac{\varpi_{D;D-2}}{2^{D}\vep^{D}}  \Gamma(D)(1-2^{-D}) \zeta_R(D+1) \mathcal{A}(D)
+\frac{1}{2\pi a_1}   \frac{\varpi_{D;D-4}}{2^{D-2}\vep^{D-2}}  \Gamma(D-2)(1-2^{2-D}) \zeta_R(D-1) \mathcal{A}(D-2)
\\&-\frac{1}{2\pi a_1}  \frac{\varpi_{D;D-2}}{2^{D-2}\vep^{D-2}}  \Gamma(D-1) (1-2^{2-D})\zeta_R(D-1) \mathcal{C}(D-2)
\\& +\frac{1}{2\pi a_1}   \frac{\varpi_{D;D-2}}{2^{D-3}\vep^{D-2 }}  \Gamma(D-2) (1-2^{4-D})\zeta_R(D-3) \mathcal{G}(D-4)+\ldots
\end{split}
\end{equation*}
If \textit{the sphere $r=a_1$ is perfectly conducting, and the sphere $r=a_2$ is infinitely permeable}, we find that  the first three leading terms of the TE and TM contributions to the Casimir free energy are given respectively by
\begin{equation*}
\begin{split}
E_{\text{TE},0}\sim  &E_{\text{TE},0}^{\text{PFA}}\Biggl\{1+\vep\frac{D-1}{2}+\vep\frac{2(D^2-4D+1)}{D(D-1)}\frac{2^D-4}{2^D-1}\frac{\zeta_R(D-1)}{\zeta_R(D+1)}\\&
+\vep^2\frac{(D-1)(3D^2-2D-17)}{24(D+2)}   +\vep^2\frac{5D^3 - 15D^2 - 59D - 15}{6D(D + 2)}\frac{2^D-4}{2^D-1}\frac{\zeta_R(D-1)}{\zeta_R(D+1)}\\
&+\vep^2\frac{2(D^4 - 6D^3 + 2D^2 + 28D - 13)}{ D(D-1)(D-2)(D + 2)}\frac{2^D-16}{2^D-1}\frac{\zeta_R(D-3)}{\zeta_R(D+1)}\Biggr\},\\
E_{\text{TM},0}\sim  &E_{\text{TM},0}^{\text{PFA}}\Biggl\{1+\vep\frac{D-1}{2}+\vep\frac{2(D^2 - 2D - 1)}{D(D - 1)}\frac{2^D-4}{2^D-1}\frac{\zeta_R(D-1)}{\zeta_R(D+1)}\\&
+\vep^2\frac{(D-1)(3D^2-2D-17)}{24(D+2)}   +\vep^2\frac{5D^3 - 3D^2 - 23D + 9}{6D(D + 2)}\frac{2^D-4}{2^D-1}\frac{\zeta_R(D-1)}{\zeta_R(D+1)}\\
&+\vep^2\frac{2(D + 1)(D^3 - 3D^2 - 3D + 11)}{(D - 1)(D - 2)D(D + 2)}\frac{2^D-16}{2^D-1}\frac{\zeta_R(D-3)}{\zeta_R(D+1)}\Biggr\},
\end{split}
\end{equation*}where $E^{\text{PFA}}_{\text{TE},0}$ and $E^{\text{PFA}}_{\text{TM},0}$ are respectively the proximity force approximations to the TE and TM contributions which are $(D-2)/(D-1)$ and $1/(D-1)$ times the total proximity force approximation \eqref{eq5_10_2}.
It follows that    the total zero temperature Casimir free energy is
\begin{equation*}
\begin{split}
E_{\text{Cas},0}\sim  &E_{\text{Cas},0}^{\text{PFA}}\Biggl\{1+\vep\frac{D-1}{2}+\vep\frac{2(D-3)}{D }\frac{2^D-4}{2^D-1}\frac{\zeta_R(D-1)}{\zeta_R(D+1)}\\&
+\vep^2\frac{(D-1)(3D^2-2D-17)}{24(D+2)}   +\vep^2\frac{5D^4 - 20D^3 - 32D^2 + 80D + 39}{6D(D-1)(D + 2)}\frac{2^D-4}{2^D-1}\frac{\zeta_R(D-1)}{\zeta_R(D+1)}\\
&+\vep^2\frac{2(D^4 - 6D^3 + 6D^2 + 24D - 37)}{ D(D-1)(D-2)(D + 2)}\frac{2^D-16}{2^D-1}\frac{\zeta_R(D-3)}{\zeta_R(D+1)}\Biggr\}.
\end{split}
\end{equation*}
When $D=3$, we have
\begin{equation}\label{eq5_6_2}
\begin{split}
E_{\text{TE},0}\sim  &\frac{7\pi^3}{2880\vep^3}\Biggl\{1+ \vep\left(1-\frac{40}{7\pi^2}\right)
+\frac{\vep^2}{15}  -\vep^2\frac{13}{7\pi^2}+\vep^2\frac{192}{ 7\pi^4}\Biggr\},\\
E_{\text{TM},0}\sim  &\frac{7\pi^3}{2880\vep^3}\Biggl\{1+ \vep\left(1+\frac{40}{7\pi^2}\right)
+\frac{\vep^2}{15}  +\vep^2\frac{27}{7\pi^2}+\vep^2\frac{192}{ 7\pi^4}\Biggr\};
\\
E_{\text{Cas},0}\sim  &\frac{7\pi^3}{1440\vep^3}\Biggl\{1+ \vep
+\vep^2\left(\frac{1}{15}  +\frac{1}{\pi^2} + \frac{192}{ 7\pi^4}\right)\Biggr\}.
\end{split}
\end{equation}

If \textit{the sphere $r=a_1$ is infinitely permeable, and the sphere $r=a_2$ is perfectly conducting}, we find that the first three leading terms of the TE and TM contributions to the Casimir free energy are given respectively by
\begin{equation*}
\begin{split}
E_{\text{TE},0}\sim  &E^{\text{PFA}}_{\text{TE},0}\Biggl\{1+\vep\frac{D-1}{2}-\vep\frac{2(D^2-4D+1)}{D(D-1)}\frac{2^D-4}{2^D-1}\frac{\zeta_R(D-1)}{\zeta_R(D+1)}\\&
+\vep^2\frac{(D-1)(3D^2-2D-17)}{24(D+2)}   -\vep^2\frac{7D^4 - 28D^3 + 8D^2 + 148D - 63}{6D(D-1)(D + 2)}\frac{2^D-4}{2^D-1}\frac{\zeta_R(D-1)}{\zeta_R(D+1)}\\
&+\vep^2\frac{2(D^4 - 6D^3 + 2D^2 + 28D - 13)}{ D(D-1)(D-2)(D + 2)}\frac{2^D-16}{2^D-1}\frac{\zeta_R(D-3)}{\zeta_R(D+1)}\Biggr\},
\end{split}
\end{equation*}\begin{equation*}
\begin{split}
E_{\text{TM},0}\sim  &E^{\text{PFA}}_{\text{TM},0}\Biggl\{1+\vep\frac{D-1}{2}-\vep\frac{2(D^2 - 2D - 1)}{D(D - 1)}\frac{2^D-4}{2^D-1}\frac{\zeta_R(D-1)}{\zeta_R(D+1)}\\&
+\vep^2\frac{(D-1)(3D^2-2D-17)}{24(D+2)}   -\vep^2\frac{7D^4 - 16D^3 - 40D^2 + 64D + 57}{6D(D - 1)(D + 2)}\frac{2^D-4}{2^D-1}\frac{\zeta_R(D-1)}{\zeta_R(D+1)}\\
&+\vep^2\frac{2(D + 1)(D^3 - 3D^2 - 3D + 11)}{(D - 1)(D - 2)D(D + 2)}\frac{2^D-16}{2^D-1}\frac{\zeta_R(D-3)}{\zeta_R(D+1)}\Biggr\}.
\end{split}
\end{equation*}
It follows that for  the total zero temperature Casimir free energy,
\begin{equation*}
\begin{split}
E_{\text{Cas},0}\sim  &E^{\text{PFA}}_{\text{Cas},0}\Biggl\{1+\vep\frac{D-1}{2}-\vep\frac{2(D-3)}{D }\frac{2^D-4}{2^D-1}\frac{\zeta_R(D-1)}{\zeta_R(D+1)}\\&
+\vep^2\frac{(D-1)(3D^2-2D-17)}{24(D+2)}   -\vep^2\frac{7D^4 - 28D^3 + 20D^2 + 112D - 183}{6D(D-1)(D + 2)}\frac{2^D-4}{2^D-1}\frac{\zeta_R(D-1)}{\zeta_R(D+1)}\\
&+\vep^2\frac{2(D^4 - 6D^3 + 6D^2 + 24D - 37)}{ D(D-1)(D-2)(D + 2)}\frac{2^D-16}{2^D-1}\frac{\zeta_R(D-3)}{\zeta_R(D+1)}\Biggr\}.
\end{split}
\end{equation*}
The $D=3$ case can be obtained by duality from \eqref{eq5_6_2} as before.

As in the high temperature region, the corrections to the proximity force approximations in the case of mixed boundary conditions are more complicated than the case of homogeneous boundary conditions. We find that the first correction terms are already different for different combinations of boundary conditions. Again, between the two scenarios of mixed boundary conditions, we find that the   force is stronger when the smaller sphere is perfectly conducting and the larger one is infinitely permeable.
\section{The low temperature asymptotic expansion of the thermal correction}
To find the low temperature asymptotics of the thermal correction, we use the Abel-Plana summation formula \cite{32,33,34}, which states that for a well-behaved function $g(z)$,
\begin{equation*}
\begin{split}
\frac{1}{2}g(0)+\sum_{p=1}^{\infty}g(p)=&\int_0^{\infty} g(x) dx +i\int_0^{\infty}\frac{g(iy)-g(-iy)}{e^{2\pi y}-1}dy  \\&+2\pi i \sum_{\text{Re}\,z\geq 0, \,\text{Im}\,z>0}w(z)\text{Res}_{z}\left\{\frac{g(z)}{e^{-2\pi i z}-1}\right\}-2\pi i \sum_{\text{Re}\,z\geq 0, \,\text{Im}\,z<0}w(z)\text{Res}_{z}\left\{\frac{g(z)}{e^{2\pi i z}-1}\right\},
\end{split}
\end{equation*}where $w(z)=1/2$ if $\text{Re}\;z=0$ and $w(z)=1$ if $\text{Re}\;z>0$.
Applying this to the Casimir free energy \eqref{eq4_21_2} gives
\begin{equation}\label{eq5_12_1}
\begin{split}
E=E_0+\frac{i}{2\pi}\sum_{l=1}^{\infty}d_l(D)\int_0^{\infty}\frac{f_l(i\xi)-f_l(-i\xi)}{e^{\frac{\xi}{T}}-1}d\xi+\text{exponentially decaying terms},
\end{split}
\end{equation}where $f_l(\xi)$ is given by \eqref{eq5_6_5}.
This formula can also be obtained by deforming the contour of integration in \eqref{eq4_21_3} from the positive real axis to the imaginary axis. The middle term in \eqref{eq5_12_1} is the term that would give the leading terms to the thermal correction in the low temperature region.
These can be obtained by expanding $f_l(i\xi)-f_l(-i\xi)$ in ascending powers of $\xi$ and applying the formula
$$\int_0^{\infty}\frac{\xi^{\mu}}{e^{\frac{\xi}{T}}-1}d\xi=\Gamma(\mu+1)\zeta_R(\mu+1)T^{\mu+1}.$$Observe that the even powers of $\xi$ in $f(\xi)$ would vanish in the expression
$f_l(i\xi)-f_l(-i\xi)$. Therefore, we only need to concentrate on the terms in $f(\xi)$ that is not even in $\xi$.
As in \cite{42}, we use the following small $z$-expansion for $I_{\nu}(z)$ and $K_{\nu}(z)$:
\begin{equation}\label{eq5_12_2}
I_{\nu}(z)=\left(\frac{z}{2}\right)^{\nu}\frac{1}{\Gamma(1+\nu )}\left(1+\mathscr{I}_{\nu}(z)\right),
\end{equation}
\begin{equation}\label{eq5_12_3}
K_{\nu}(z)=\left\{\begin{aligned}&\frac{\pi}{2}(-1)^{l+\frac{D-3}{2}}\left(\frac{z}{2}\right)^{-\nu}\frac{1}{\Gamma(1-\nu)}\left( 1+\mathscr{I}_{-\nu}(z)
-\left(\frac{z}{2}\right)^{2l+D-2} \frac{\Gamma(1-\nu)}{\Gamma(1+\nu)}\left[ 1+\mathscr{I}_{\nu}(z)\right]\right),\quad D\quad\text{odd}\\
 &\frac{1}{2}\left(\frac{z}{2}\right)^{-\nu}\Gamma(\nu)\left(1+\mathscr{J}_{\nu}(z)+2(-1)^{\nu+1}\left(\frac{z}{2}\right)^{2l+D-2}\frac{1}{\Gamma(\nu)\Gamma(\nu+1)}\left[1+
\mathscr{I}_{\nu}(z)\right]\ln z\right),\quad   D\quad\text{even}.\end{aligned}\right.
\end{equation}Here $\mathscr{I}_{\nu}(z)$ and $\mathscr{J}_{\nu}(z)$ are functions that only contain positive even powers of $z$.
From \eqref{eq5_12_2} and \eqref{eq5_12_3}, it follows that
\begin{equation*}
\begin{split}
i(f_l(i\xi)-f_l(-i\xi))\sim 2\pi\mathscr{A}_l\xi^{2\nu},
\end{split}
\end{equation*}
where
\begin{equation*}
\begin{split}
\mathscr{A}_l=&\left[
 \frac{1}{\nu\Gamma(\nu)^2} \frac{1}{2^{2\nu}}\left(\frac{\alpha_1+\beta_1\nu}{\alpha_1-\beta_1\nu}a_1^{2\nu}
-\frac{\alpha_2+\beta_2\nu}{\alpha_2-\beta_2\nu} a_2^{2\nu}\right)\right]\left/
\left[\frac{\alpha_1-\beta_1\nu}{\alpha_1+\beta_1\nu}\frac{\alpha_2+\beta_2\nu}{\alpha_2-\beta_2\nu}\left(\frac{a_2}{a_1}\right)^{2\nu}-1\right]\right.\\
=&-\frac{1}{\nu\Gamma(\nu)^2} \frac{1}{2^{2\nu}} \frac{\alpha_1+\beta_1\nu}{\alpha_1-\beta_1\nu}a_1^{2\nu}.
\end{split}
\end{equation*}From these, we can see that the leading order term of $f_l(i\xi)-f_l(i\xi)$ is of order $\xi^{2l+D-2}$. Therefore, when $d T\ll a_2T\ll 1$, the leading thermal correction comes from the term with $l=1$. This implies that
\begin{equation*}
\begin{split}
\Delta_T E\sim & -\frac{d_1(D)}{D\Gamma\left(\frac{D}{2}\right)^2} \frac{1}{2^{D-1}}\frac{2\alpha_1+\beta_1D}{2\alpha_1-\beta_1D}a_1^{D}\Gamma(D+1)\zeta_R(D+1)T^{D+1}+\ldots\\
=&-\frac{d_1(D)}{\sqrt{\pi}a_1}\frac{\Gamma\left(\frac{D+1}{2}\right)}{\Gamma\left(\frac{D}{2}\right)}\frac{2\alpha_1+\beta_1D}{2\alpha_1-\beta_1D}\zeta_R(D+1)(aT)^{D+1}+\ldots.
\end{split}
\end{equation*} Notice that the leading term in the thermal correction does not depend on the boundary conditions and the radius of the larger sphere.
Using the fact that $$b_1(D)=D,\hspace{1cm}h_1(D)=\frac{D(D-1)}{2},$$ we find that if the smaller sphere is perfectly conducting,
then the leading terms of the thermal corrections of the TE contribution, the TM contribution and the total Casimir free energy are given respectively by
\begin{equation}\label{eq5_12_5}
\begin{split}
\Delta_T E_{\text{TE}}\sim &  -\frac{D(D-1)}{2\sqrt{\pi}a_1}\frac{\Gamma\left(\frac{D+1}{2}\right)}{\Gamma\left(\frac{D}{2}\right)} \zeta_R(D+1)(a_1T)^{D+1}+\ldots,
\\
\Delta_T E_{\text{TM}}\sim &  \frac{D(D-1)}{\sqrt{\pi}a_1}\frac{\Gamma\left(\frac{D+1}{2}\right)}{\Gamma\left(\frac{D}{2}\right)} \zeta_R(D+1)(a_1T)^{D+1}+\ldots,\\
\Delta_T E_{\text{Cas}}\sim &  \frac{D(D-1)}{2\sqrt{\pi}a_1}\frac{\Gamma\left(\frac{D+1}{2}\right)}{\Gamma\left(\frac{D}{2}\right)} \zeta_R(D+1)(a_1T)^{D+1}+\ldots.
\end{split}
\end{equation}
Note that the leading term of the TM contribution is always negative twice the leading term of the TE contribution.
If the smaller sphere is infinitely permeable,
then
\begin{equation}\label{eq5_12_6}
\begin{split}
\Delta_T E_{\text{TE}}\sim &  \frac{D(D-1)}{(D-2)\sqrt{\pi}a_1}\frac{\Gamma\left(\frac{D+1}{2}\right)}{\Gamma\left(\frac{D}{2}\right)} \zeta_R(D+1)(a_1T)^{D+1}+\ldots,\\
\Delta_T E_{\text{TM}}\sim &  -\frac{D}{\sqrt{\pi}a_1}\frac{\Gamma\left(\frac{D+1}{2}\right)}{\Gamma\left(\frac{D}{2}\right)} \zeta_R(D+1)(a_1T)^{D+1}+\ldots,\\
\Delta_T E_{\text{Cas}}\sim &  \frac{D }{(D-2)\sqrt{\pi}a_1}\frac{\Gamma\left(\frac{D+1}{2}\right)}{\Gamma\left(\frac{D}{2}\right)} \zeta_R(D+1)(a_1T)^{D+1}+\ldots.
\end{split}
\end{equation}When $D\geq 4$, we notice that the thermal correction is larger when the smaller sphere is perfectly conducting.
When $D=3$, we have specifically
\begin{equation}\label{eq5_12_11}\Delta_T E_{\text{Cas}}\sim  \frac{\pi^3}{15}a_1^3T^4+\ldots\end{equation} for any boundary conditions.
 Notice that this is the negative of the low temperature leading term of the thermal correction to the Casimir free energy of a single perfectly conducting sphere of radius $a_1$ \cite{4,1}. In fact, one can show that  \eqref{eq5_12_5} and \eqref{eq5_12_6} give respectively the negative of the leading thermal correction for a single perfectly conducting sphere and a single infinitely permeable sphere in $D$-dimensional space.

In the case of two   infinite parallel plates, the low temperature leading term of the thermal correction to the   force density is given by \cite{68}:
$$\Delta_T\mathcal{F}_{\text{Cas}}^{\parallel}=-(D-1)\frac{\Gamma\left(\frac{D+1}{2}\right)}{\pi^{\frac{D+1}{2}}}\zeta_R(D+1)T^{D+1},$$regardless of the boundary conditions on the plates.
Multiplying by the area of the sphere of radius $a_1$,   the proximity force approximation of the low temperature leading term of the thermal correction is
\begin{equation}\label{eq5_12_7}\Delta_TF=-\frac{2(D-1)}{\sqrt{\pi}a_1^2}
\frac{\Gamma\left(\frac{D+1}{2}\right)}{\Gamma\left(\frac{D}{2}\right)}\zeta_R(D+1)(a_1T)^{D+1}.\end{equation}
However, the leading terms \eqref{eq5_12_5} and \eqref{eq5_12_6} derived from the exact formulas  of the Casimir free energies showed that when the smaller sphere is perfectly conducting, then the low temperature leading term of the thermal correction to the    force is
\begin{equation}\label{eq5_12_8}\Delta_TF_{\text{Cas}}\sim  -\frac{D^2(D-1)}{2\sqrt{\pi}a_1^2}\frac{\Gamma\left(\frac{D+1}{2}\right)}{\Gamma\left(\frac{D}{2}\right)} \zeta_R(D+1)(a_1T)^{D+1}+\ldots.\end{equation} Whereas if the smaller sphere is infinitely permeable, then the low temperature leading term of the thermal correction to  the   force is
\begin{equation}\label{eq5_12_9}\Delta_T F_{\text{Cas}}\sim -  \frac{D^2 }{(D-2)\sqrt{\pi}a_1^2}\frac{\Gamma\left(\frac{D+1}{2}\right)}{\Gamma\left(\frac{D}{2}\right)} \zeta_R(D+1)(a_1T)^{D+1}+\ldots.\end{equation}
As expected, these do not agree  with  the temperature correction in the proximity force approximation \eqref{eq5_12_7}. The point is that under the low temperature condition $dT\ll a_2T\ll 1$, the thermal correction is much smaller than the zero temperature Casimir energy by an order $\varepsilon^D$ and, thus, the proximity force approximation is not applicable. However, the proximity force approximation remains applicable for the calculation of the total free energy as we have seen in Section \ref{sl}. Similar situation has been observed in $D=3$ dimensions for the case of a sphere in front of a plate \cite{42}.

\section{  Conclusion}
 In this article, we studied the Casimir interaction between two concentric spheres in $(D+1)$-dimensional spacetime due to the confinement of the electromagnetic field between the spheres. We consider the cases of perfectly conducting -- perfectly conducting, infinitely permeable -- infinitely permeable, perfectly conducting -- infinitely permeable and infinitely permeable -- perfectly conducting boundary conditions on the spheres. The first two are referred to as homogeneous boundary conditions, and the last two are called mixed boundary conditions. For homogeneous boundary conditions, the Casimir interaction between the spheres is always attractive. For mixed boundary conditions, it is always repulsive.

 We are particularly interested in studying the asymptotic behaviors of the Casimir free energy when $\vep$, the ratio of the separation between the spheres to the radius of the smaller sphere, is small. Both the high temperature region and the low temperature region are considered. In the high temperature region, the Casimir free energy is dominated by the classical term which is the term corresponding to the zeroth Matsubara frequency. In the case of two concentric spheres, this term is quite simple. It can be written as a series in elementary functions. In the low temperature region, the Casimir free energy is dominated by the zero temperature term, which has to be expressed in terms of Bessel functions.
The first three leading terms are computed explicitly. In the high temperature region, the leading terms are of order $T\vep^{1-D}$, and they coincide with that obtained using proximity force approximation. For the zero temperature terms, the leading terms are of order $\vep^{-D}$, and they also agree with the proximity force approximations. It is interesting to observe that  the asymptotic expansions of the Casimir free energies have the following universal structure:
 \begin{equation*}
 E_{\text{Cas}}=E_{\text{Cas}}^{\text{PFA}}\left(1+\vep\frac{D-1}{2}+\vep \mathscr{B}_0 (D)\frac{\zeta(D+i-2)}{\zeta(D+i)}+\vep^2 \left[\mathscr{C}_2(D)+\mathscr{C}_1(D)\frac{\zeta(D+i-2)}{\zeta(D+i)}+\mathscr{C}_0(D)\frac{\zeta(D+i-4)}{\zeta(D+i)}\right]+\ldots\right),
 \end{equation*}where $i=1$ in the low temperature region, and $i=0$ in the high temperature region.   $\mathscr{B}_0(D),\mathscr{C}_0(D),\mathscr{C}_1(D),\mathscr{C}_2(D)$ are rational functions of $D$ that are $O(D^0), O(D^0), O(D), O(D^2)$ when $D$ is large. In the case of homogeneous boundary conditions, the terms $\mathscr{B}_0 (D)$ and $\mathscr{C}_0 (D)$ are absent.

In general, the corrections to the proximity force approximations are more complicated in the case of mixed boundary conditions compared to the case of homogeneous boundary conditions. In fact, for homogeneous boundary conditions, the first correction is the same when the two spheres are both perfectly conducting or both infinitely permeable. For  the two scenarios of mixed boundary conditions, the first corrections are different. It is observed that the Casimir interaction is stronger when the smaller sphere is perfectly conducting and the larger sphere is infinitely permeable.

 Finally, the low temperature leading terms of the thermal corrections to the Casimir free energies are computed. They are finite when $\vep\rightarrow 0^+$ and are of order $T^{D+1}$. It is interesting to find that these leading terms are independent of the larger sphere. They do not depend on the radius or the boundary conditions on the larger sphere. As have been observed  by a few researchers for other geometric configurations \cite{42,67}, this case is outside the application region of the proximity force approximation.

For future works, it would be interesting  to consider eccentric spheres and compare the results with those obtained here.

\vspace{1cm}
\begin{acknowledgments}
We have benefited from discussions with K. Kirsten and A. Flachi. This project is funded by Ministry of Higher Education of Malaysia   under FRGS grant FRGS/2/2010/SG/UNIM/02/2. We also appreciate the helpful comments given by the anonymous referee.
\end{acknowledgments}

\end{document}